\documentclass[showpacs,floatfix,amsmath,superscriptaddress,eqsecnum,amssymb,prb,twocolumn]{revtex4}
\usepackage{latexsym}
\usepackage{graphicx}
\usepackage{psfrag}

\newcommand{\be}{\begin{equation}}
\newcommand{\ee}{\end{equation}}
\newcommand{\ba}{\begin{eqnarray}}
\newcommand{\ea}{\end{eqnarray}}
\newcommand{\baa}{\begin{eqnarray*}}
\newcommand{\eaa}{\end{eqnarray*}}

\def\prl#1#2#3{Phys.\ Rev.\ Lett.\ {\bf #1}, #2 (#3)}

\def\prb#1#2#3{Phys.\ Rev.\ B {\bf #1}, #2 (#3)}

\def\rmp#1#2#3{Rev.\ Mod.\ Phys.\ {\bf #1}, #2 (#3)}

\def\physicac#1#2#3{Physica C {\bf #1}, #2 (#3)}

\def\nature#1#2#3{Nature {\bf #1}, #2 (#3)}
\def\science#1#2#3{Science {\bf #1}, #2 (#3)}

\def\pnas#1#2#3{Proc. Natl. Acad. Sci. U.S.A. {\bf #1}, #2 (#3)}

\def\rpp#1#2#3{Rep. Prog. Phys.\ {\bf #1}, #2 (#3)}

\def\be{\begin{equation}}
\def\ee{\end{equation}}
\def\ba{\begin{eqnarray}}
\def\ea{\end{eqnarray}}

\def\LSCO{La$_{2-x}$Sr$_x$CuO$_4$}
\def\LCO{La$_2$CuO$_4$}

\def\YBCO{YBa$_2$Cu$_3$O$_{6+y}$}

\def\BSCCO{Bi$_2$Sr$_2$CaCu$_2$O$_{8+\delta}$}
\def\C60{A$_x$C$_{60}$}
\def\oxychloride{Ca$_{2-x}$Na$_x$CuO$_2$Cl$_2$}

\def\HgCu3{HgCa$_2$Cu$_3$O$_{8+y}$}
\def\HgCu4{HgBa$_2$Ca$_3$Cu$_4$O$_{10+y}$}
\def\TlCu{Tl$_2$Ba$_2$CuO$_{6+\delta}$}
\def\TlCu3{Tl$_2$Ba$_2$Ca$_2$Cu$_3$O$_{10+y}$}
\def\TlCu4{Tl$_2$Ba$_2$Ca$_3$Cu$_4$O$_{12+y}$}

\def\BiCu3{Bi$_2$Sr$_2$Ca$_{2}$Cu$_3$O$_y$}

\def\8LSCO{La$_{1.88}$Sr$_{.12}$CuO$_4$}
\def\110LNSCO{La$_{1.5}$Nd$_{0.4}$Sr$_{0.1}$CuO$_{4}$}
\def\stage4LCO{La$_{2}$CuO$_{4+\delta}$}
\def\Y248{YBa$_2$Cu$_4$O$_8$}

\def\NbSe2{NbSe$_2$}
\def\TaSe2{TaSe$_2$}
\def\TiSe2{TiSe$_2$}

\begin{document}

\title{Surface pinning of fluctuating charge order: an ``extraordinary" 
surface phase transition. }
\author{Stuart E. Brown}
\affiliation{Department of Physics and Astronomy,
UCLA, Los Angeles, CA 90095-1547}
\author{Eduardo Fradkin}
\affiliation{Department of Physics, 
University of Illinois, 1110  W. Green St., Urbana, Illinois  61801-3080, USA} 
\author{Steven A. Kivelson}
\affiliation{Department of Physics and Astronomy,
Stanford University, Stanford CA 94305}
\affiliation{Department of Physics and Astronomy,
UCLA, Los Angeles, CA 90095-1547}

\date{\today}

\begin{abstract}
We study the mean-field theory of charge-density wave (CDW) order in a layered 
system, including the effect of the long-range Coulomb interaction and of 
screening by uncondensed electrons.  We particularly focus on the conditions 
necessary for an ``extraordinary'' transition, in which the surface orders at 
a higher temperature, and is more likely to be commensurate, than the bulk.  
We interpret recent experiments on {\oxychloride} as indicating the presence 
of commensurate CDW at the surface that is {\it not} present in the bulk.  
More generally, we show that poor screening of the Coulomb interaction tends 
to stabilize incommensurate order, possibly explaining why the CDW order in 
{\LSCO} and {\NbSe2} remains incommensurate to $T\to 0$, despite the small 
magnitude of the incommensurability.

\end{abstract}

\maketitle

With the advent of high resolution angle resolved photoemission (ARPES) and 
scanning tunneling spectroscopy (STS), there is increasing interest in looking 
for evidence of novel order or incipient order in strongly correlated electron 
systems by studying the electronic structure of the surface layer  
-- some of the most interesting recent evidence that charge order plays a 
critical role in the physics of the cuprate high temperature superconductors 
(HTC) comes from such 
studies\cite{davis-vortex-science-02,howald-prb-03,ali-science-04,
hanaguri-davis-oxychloride-04}. 
A persistent question about such studies arises, 
``Is the surface electronic structure the same as the bulk?'' 

In the present note, we outline some of the possibilities for transitions to 
ordered states {\em at the surface} of a bulk system, where the surface order 
reflects, but in somewhat subtle and indirect ways, the character of the bulk. 
In particular, we argue that the ``commensurate checkerboard order'' recently 
discovered\cite{hanaguri-davis-oxychloride-04} in \oxychloride\ 
(NaCCOC) is very likely  {\it not} directly representative of charge order in 
the bulk of the sample. However, we describe suggestive, but not conclusive 
reasons to believe that this surface order {\it is} a pinned relative of 
fluctuating charge order in the bulk 
- probably related to the fluctuating 
charge stripe order seen in bulk measurements on 
\LSCO (LSCO). This work extends earlier investigations by two of us\cite{rmp} 
of general strategies for observing correlations - ``fluctuating order'' - 
which reflect the proximity in a generalized phase diagram of a true ordered state. 

When a bulk system undergoes a phase transition to a broken symmetry state, 
such as a charge-density wave (CDW) state, the surface of the system must reflect 
the broken symmetry, as well. However, 
%since fluctuation effects are stronger at the surface, 
%and ordering temperatures lower as there are fewer 
%neighbors, 
one might expect order to be weaker at the surface.
%, so that there might be a range of temperatures $T$ in which the bulk 
%is ordered while the surface is 
%partially melted.
% (For instance, there could be an intermediate range of 
%temperatures for which the bulk CDW is 
%commensurate, while it is incommensurate at the surface.)
 Nonetheless, there are known cases of {\em surface phase transitions} 
 in which an ``extraordinary transition'' occurs\cite{extraordinary}, 
 {\it i.e.\/} a phase transition in which  the surface orders at a higher 
 temperature than the bulk. 

In the present paper, we consider the case of a CDW in a layered (quasi 2D) material with a smooth surface obtained by cleaving between two layers, for which: \\

%\begin{itemize}
%\item{ 
1) We analyze the circumstances under which an extraordinary surface phase 
transition can occur. We show that if the couplings within each layer, 
including the surface layer, are identical, the surface is unlikely to 
order before the bulk. However,  phonon modes associated with the motion 
of atoms transverse to the layers tend to be softer at the surface than 
in the bulk.  If the coupling to such modes is sufficiently strong, an 
extraordinary surface phase transition occurs. 
An example of such a phonon mode is the apical O or Cl modes  in LSCO and NaCCOC, 
respectively, which independent studies suggest are strongly coupled to 
the charge density in the Cu-O planes. 
%We also discuss the role of Coulomb interactions and of the existence 
%of a compressible conducting fluid 
%(``quasiparticles") coexisting with bulk CDW order.
\\ 

\begin{figure}
\psfrag{Ti}{$T_I$}
\psfrag{Tsi}{$T_{sI}$}
\psfrag{Tc}{$T_C$}
\psfrag{Tsc}{$T_{sC}$}
\psfrag{BD}{ $BD$}
\psfrag{SD}{$ SD$}
\psfrag{BI}{$BI$}
\psfrag{BC}{$BC$}
\psfrag{SI}{$SI$}
\psfrag{SC}{$SC$}
\psfrag{T}{$T$}
\psfrag{r}{$r_0/r$}
\psfrag{1}{$1$}
\psfrag{D}{\small Bulk \;\;\; Disordered}
\psfrag{C}{\small Bulk \;\;\; Commensurate}
\psfrag{I}{\small Bulk \;\;\;  Incommensurate}
\begin{center}
\includegraphics[width=0.4 \textwidth]{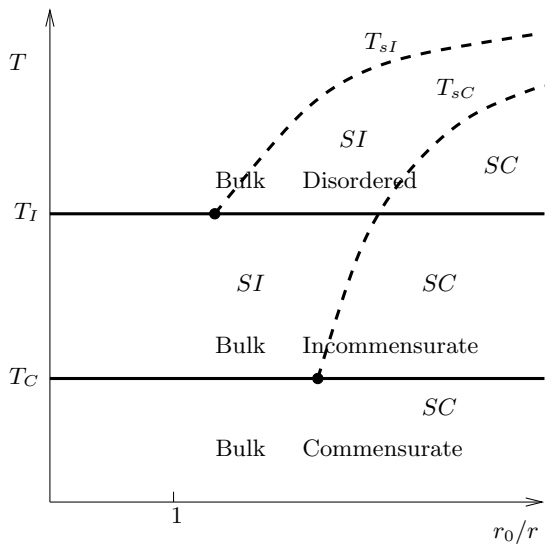}
\end{center}
\caption{Schematic one-parameter cut of the phase diagram for a CDW system with a surface: $r$ and
$r_0$ measure the strength of the quadratic term in the McMillan free energy 
in the bulk and on the
surface. The full and broken lines represent the bulk and surface phase transitions respectively: $T_I$ and $T_C$ are the bulk critical temperatures for the
disordered-incommensurate and for the bulk incommensurate-commensurate
phase transitions; $T_{sI}$ and $T_{sC}$ are the corresponding surface
phase transitions. The extraordinary surface transitions for $r_0/r$ are shown. SI: surface incommensurate CDW; SC: surface commensurate CDW. Other surface orderings ({\it e.g.\/} ``ordinary''surface transitions) are possible but are not shown. }
\label{fig:phase_diagram}
\end{figure}

%\item{
2) Under circumstances in which an anomalous transition occurs, one expects from 
simple Landau-Ginzburg considerations, the following hierarchy of transition 
temperatures: $T_{sI} \ge T_{sC}$ and $T_{sI} \ge T_{I}$, 
where $ T_{sI}$, $T_{sC}$, and $T_{I}$ are the transition temperatures at 
which surface incommensurate CDW order, surface commensurate CDW order, 
and bulk incommensurate order onsets, respectively. 
Moreover, as is generally the case, $T_{I} \ge T_{C}$, where $T_{C}$ 
marks the bulk transition to a commensurate CDW. 
The extraordinary transition is particularly dramatic when $T_{C}$ and 
even $T_{I}$ are zero, {\it i.e.\/} when the bulk system is in a quantum 
disordered phase, but possibly with a commensurate CDW on the surface layer.  
This sequence of transitions is shown schematically in 
Fig. \ref{fig:phase_diagram}. \\
%%%%%%%%%%%%%%%%%%%%%%%%%%%%%%%%%%%%%

%\item{
3) We interpret the STS experiments\cite{hanaguri-davis-oxychloride-04} 
on NaCCOC as being indicative of a commensurate CDW phase on the surface 
(although the large effect of quenched disorder apparent in the STS images 
make this conclusion far from certain). We show that bulk measurements 
appear  inconsistent with the existence of a commensurate CDW in the bulk.
%, and even with the existence of bulk incommensurate order of any 
%great magnitude. 
Thus, it is possible that either the order on the surface
%is entirely a consequence of surface reconstruction, with 
has no relation to the electronic properties of the bulk, or  that 
there is an extraordinary surface phase transition which reflects 
the existence of bulk electronic correlations corresponding to a 
nearly ordered CDW in the bulk.  By comparing the character of the 
observed surface order with modulations with similar periodicity 
(but rather different character) seen in STS studies of \BSCCO (BSCCO)
surfaces\cite{davis-vortex-science-02,howald-prb-03,ali-science-04}, 
and with stripe order and fluctuating order seen in bulk diffraction 
studies of LSCO and \YBCO (YBCO), we tentatively favor the latter 
interpretation. \\
%\end{itemize}

4)  In the course of this study, we have been forced to study the non-local 
Coulomb interaction.  Under most circumstances -- certainly, at any finite 
temperature or when the Fermi surface is incompletely gapped by the CDW 
-- the Coulomb interactions between CDW fluctuations are screened, with a 
screening length $\xi$.  While when $\xi$ is small, the usual Landau-Ginzburg 
theory is recovered, we have found that poor screening (large $\xi$) 
substantially stabilizes the incommensurate CDW 
-- it both tends to increase $T_I$ and decrease $T_C$.  
It is possible that this explains the remarkable stability of 
the incommensurate phase in some systems, despite very small values 
of the incommensurability.  For instance, the ordering wave vector in 
{\NbSe2} is roughly 1\% incommensurate, but there is apparently no 
transition to a commensurate state even in the limit $T\to 0$.\\

The paper is organized as follows.  
In Section \ref{sec:landau} we introduce a Landau-Ginzburg theory of 
quasi-2D CDW order, including the effects of Coulomb interactions and
screening on the bulk CDW phase transitions. 
In Section \ref{sec:extra} we discuss the extraordinary surface phase 
transition. In Section \ref{sec:phonons} we discuss the role of 
surface phonons in the ``mechanism.''  
In Section \ref{sec:NaCCOC} we discuss the case of NaCCOC and other cuprates.  
Readers who are exclusively interested in the application of these ideas 
to the experimental system can skip directly to this section.

\section{Landau-Ginzburg Model of quasi 2D CDW order}
\label{sec:landau}

To discuss these issues in the context of an explicit model, 
we consider the case of a CDW in a quasi 2D (layered) system with 
tetragonal symmetry\cite{subir-new}.  
We consider the case in which the local considerations in each plane 
favor density wave order with two, mutual orthogonal ordering vectors, 
$\vec Q_1$ and $\vec Q_2$, ($| \vec Q_1 | = |\vec Q_2|$ since they are 
related by rotation by $\pi/2$) which lie along a preferred symmetry axis of the 
crystal.  We then express the theory in terms of two complex order parameter fields, 
$\psi_{1,n}$ and $\psi_{2,n}$, such that the local charge density $\rho_n(\vec x)$ 
in plane $n$ is
\ba
&&\rho_n(\vec x)  =\rho_n^N(\vec x) + \nonumber\\
&&
\sum_j \left[ i \Lambda \;  \psi_{j,n}^*(\vec x) \hat Q_j \cdot  \vec\nabla \psi_{j,n}(\vec x) +
 \psi_{j,n}(\vec  x) \; e^{i \vec  Q_j \cdot \vec x }  + \rm{c.c.}\right] 
 \nonumber \\
&&+{\rm higher\;  harmonics} %\nonumber
 \label{density}
\ea
where $\rho_n^N$ is the ``normal''  component of the charge density, {\it i.e.\/} 
the part which is not tied to the CDW, and $\hat Q_j= \vec Q_j/|\vec Q_j|$ 
are two unit vectors along the two CDW ordering directions.  
The terms represented as ``higher harmonics" refer to components of the 
density at higher harmonics of the fundamental periods, $n_1\vec Q_1+n_2\vec Q_2$ with 
$|n_1|+|n_2| > 1$;  we will simply assume that these harmonics can be integrated out, 
and their effects captured by non-linear couplings to the fundamentals.  The dimensionless 
%coupling constant 
parameter $\Lambda$ reflects the change of the density which results from a compression 
of the CDW.

We are going to consider the case in which the CDW is commensurate or weakly incommensurate, 
and with the situation relevant to NaCCOC in mind, we have considered the case in which the 
CDW is near commensurability 4.   
In this case, it is convenient to take $4\vec Q_j=\vec G_j$ 
where $\vec G_j$ is a reciprocal lattice vector. Hence, a uniform amplitude 
for the CDW order parameter $\psi = {\rm const.}$ corresponds to a {\em commensurate} 
CDW while
$\psi \propto \exp[i\delta x]$ corresponds to an {\em incommensurate} CDW, where $\delta$ 
(which will be derived by minimizing the effective Hamiltonian) 
is the ``incommensurability.''  
The  case in which the CDW is far from being commensurate involves no new physics, 
so we will not discuss it explicitly.

In terms of the order parameters $\psi_1$ and $\psi_2$, we can write a McMillan 
(Landau-Ginzburg) type effective free energy functional
\cite{mcmillan-cdw-lg-75,mcmillan-discommensurations-76,bak-82} suitable for a CDW 
in a layered system (keeping lowest order terms in gradients and all terms allowed 
by symmetry through order $\psi^4$):
\ba
&&{\mathcal F}=\sum_n\int \ d^2x \left\{{\mathcal L}_{0}+
{\mathcal L}_1+{\mathcal L}_{\rm c}\right\}+{\mathcal F}_{\rm coul}+F_N 
\nonumber\\
 &&{\mathcal L}_{0}=\frac{1}{2} 
 \sum_j\left\{K |(i\hat Q_j\cdot\vec \nabla+\delta_0) \psi_{j,n}(\vec x)|^2 \right. 
 \nonumber \\
 &&\;\;\;\;\;\;\;\;\;\;\;\;\;\;\; \left.
 +K^{\prime} |\hat Q_j\times \vec \nabla \psi_{j,n}(\vec x)|^2 + 
 r \; |\psi_{j,n}(\vec x)|^2\right\} \nonumber \\
 &&{\mathcal L}_1= u ( |\psi_{1,n}(\vec x)|^2+|\psi_{2,n}(\vec x)|^2)^2 + 
 \gamma|\psi_{1,n}(\vec x)|^2|\psi_{2,n}(\vec x)|^2
 \nonumber \\
 &&{\mathcal L}_{\rm c}=V_{\rm c}\sum_{j} \left\{ \psi_{j,n}(\vec x)^4  + 
 {\rm c.c.}\right\} \nonumber \\
&& {\mathcal F}_{\rm coul}= \frac {e^2} {2\epsilon} 
\sum_{n,n^{\prime}}\int d^2 x \ d^2 x^{\prime}\frac {[ \rho_n(\vec x)-\bar \rho]
 [ \rho_{n^{\prime}}(\vec x^{\; \prime})-\bar \rho]}{\sqrt{(\vec x-\vec x^{\; \prime})^2+
 a^2(n-n^{\prime})^2}}     \nonumber\\
 &&
 \label{LGW}
 \ea
 where $n=0,1,\ldots$ labels the layers with $n=0$ being the surface layer of a 
 semi-infinite system; $j=1,2$ labels the two CDW order parameters.
In Eq.\eqref{LGW}, $r$, $u$ and $\gamma$ are phenomenological couplings which depend 
weakly on the temperature $T$. 
As usual, $r$ changes from positive to negative with decreasing temperature, 
and so is the one parameter whose temperature dependence will be explicitly considered, 
$r =\alpha_0(T-T_I^0)$.  Moreover, near the surface,  the various parameters could also 
depend on the layer index, $n$.  For simplicity, we will assume that only the surface 
layer ($n=0$) is distinct, and that the most important difference between the surface 
layer and the bulk is an additive correction to $r_0=r+\delta r = \alpha_0(T-T^0_{sI})$, 
{\it i.e.\/} a distinct mean-field transition at the surface.
%we have chosen units of length such that the compressional stiffness of the CDW is 1, 
%and scale $\psi$ %so that the first quartic term has coefficient 1.  

In Eq.\eqref{LGW} we have assumed that we can neglect all inter-plane interactions except 
the Coulomb coupling, ${\mathcal F}_{\rm coul}$. In Eq.\eqref{LGW}  $\bar\rho$ is a uniform 
background charge density, $K$ and $K^\prime$ are the CDW stiffnesses, 
and $a$ is the lattice spacing between planes.    Here, the shear stiffness, 
$K^{\prime}$ will play little role in the present discussion.  
The sign of $\gamma$ determines whether the ordered state is a unidirectional CDW, 
$\gamma > 0$, or an isotropic checkerboard, $\gamma < 0$. 
To simplify explicit expressions, we will assume that $|\gamma|$ is small, although 
so long as $\gamma > -2u$ (necessary for stability). No qualitative results depend on 
this assumption. Finally, ${\mathcal L}_{\rm c}$ is the lock-in potential which favors 
a (period $4$) CDW, $V_{\rm c}$ is the strength of the commensurability, 
and $e^2/\varepsilon$ is the strength of the Coulomb interaction.
% with $\varepsilon$ being the dielectric constant of the material.

We introduced a term in Eq.\eqref{LGW} , $F_N$, %assume that there is another term 
%in the effective action 
which governs the fluctuations of the normal density;  for our purposes, 
what is important is the preferred density, $\bar\rho^N$, and the small 
fluctuations about it (which lead to screening), so we take
\be
F_N=\sum_{n=0}^\infty \int\ d^2x \frac {\kappa_0} 2 [\rho_n^N(\vec x) - \bar\rho^N]^2
\ee
where $\kappa_0$ is the inverse compressibility of the normal fluid. 
Notice that the normal density and the CDW order parameters are coupled 
through the Coulomb interaction in the Landau-Ginzburg effective theory, 
as given by Eq.\eqref{density} and Eq.\eqref{LGW}. Integrating out the 
fluctuations of the normal excitations, parametrized by $\rho_n^N(\vec x)$, 
result in an effectively screened Coulomb interaction with a screening length 
$\xi=(4\pi e^2/\varepsilon \kappa_0)^{-1/2}$, and  with the density expressed 
as in Eq.\eqref{density} with the replacement $\rho_n^N(\vec r) \to \bar\rho^N$.

Before we proceed further, it is worth commenting on the ways in with the 
present free energy differs from the usual Landau-Ginzburg treatment 
(obtained by taking the limit $\kappa_0\to 0$ in the above), 
which does not treat the effects of the non-local Coulomb interactions.  
For $\kappa_0=0$, the incommensurability is determined directly by 
$\delta=\delta_0$, and so has no interesting temperature dependence 
that is not put in by hand, and the mean-field transition to the 
incommensurate state occurs just where $r$ changes sign, $T_I=T_I^0$, 
or at the surface at $T_{sI}=T_{sI}^0$.  For imperfectly screened Coulomb 
interactions, however, the situation is more complex.  The optimal value 
of $\delta$ is determined by a combination of $\delta_0$ and 
$\Delta \bar \rho \equiv \bar \rho-\bar\rho^N$.  
In  the limit of no screening, $\kappa_0\to\infty$, the incommensurate state 
is stable at all temperatures ($T_I\to \infty$), and at long distances, the 
incommensurability is set by the {\it constraint}, 
$\delta=-\Delta\bar\rho[2\Lambda|\psi|^2]^{-1}$.  
For intermediate values of the screening, the behavior is intermediate 
between these two limits.  Finally, since the Coulomb interaction couples 
different planes, the surface transition temperature is not determined 
solely by the values of the parameters in the surface layer, but depends 
on the coupling between the surface and the bulk in a non-trivial fashion.

Since our principal focus is on the possibility of surface transitions, we will 
treat the bulk properties cursorily.  Just to be specific, we consider the case 
of a striped phase ($\gamma >0$) in which $\psi_2=0$, but the generalization to 
checkerboard phases, which occur for $\gamma<0$,  is straightforward.  In order 
to determine the bulk (mean field) phase diagram we compare the energies of  
three different forms of the order parameter to get a sense of the bulk phase diagram:\\  
{\noindent 1)}  The disordered state, $\psi_{1,n}=0$.  The mean-field free energy 
density of the disordered state is 
\be
F_{\rm dis}=(1/2) \kappa_0\; (\Delta\bar\rho)^2.
\ee
2)  The commensurate solution,  $\psi_{1,n}(\vec r) = \psi$.  
The free energy of the commensurate state is 
    \ba
    F_{\rm c}=F_{\rm dis}+(1/2)\;[r + 
    K\delta_0^2 V_Q] |\psi|^2+[u-2 V_{\rm c}]|\psi_{j}|^{4}, \nonumber
    \ea
   where $V_Q= 4\pi e^2/(\epsilon[Q^2+\xi^{-2}])$.
\\
\noindent
      3)  The harmonic incommensurate state, 
      $\psi_{j,n}(\vec r) = \psi_j\; \exp[i\delta\hat Q_j\cdot \vec r]$. 
      (We know from the work of McMillan\cite{mcmillan-discommensurations-76} 
      that, especially near the commensurate to incommensurate transition, 
      the structure of the incommensurate state is highly anharmonic, 
      and this anharmonicity has a significant quantitative effect on the phase diagram, 
      but one can understand much of the qualitative physics ignoring this.)  
      The optimal incommensurability and free energy of this state are, respectively,
\ba
\delta=&&\delta_0[1 + A^{-1}] \ \left[1+2A^{-1}|\tilde\psi|^2\right]^{-1} \\
F_I= &&F_{\rm dis} + \frac {r+V_Q} 2 |\psi|^2+u|\psi|^4\nonumber \\
&&-\frac{\kappa_0(\Delta\bar\rho)^2}{2}\; 
\left[\frac{1+2A -|\tilde\psi|^2}{1 + A|\tilde \psi|^2} \right]\; |\tilde\psi|^2
\nonumber \\
&&
\label{FI}
\ea 
where $\tilde\psi=[2\Lambda\delta_0/\Delta\bar\rho]\psi$ and 
$A=K\delta_0/2\kappa_0\Lambda\Delta\bar\rho$.  

One interesting consequence of these expressions is that they imply 
a non-trivial temperature dependence of the incommensurability as the 
order parameter grows.  This is a general feature of an incommensurate 
state, but what is new here is the singular temperature dependence 
inherited from the $T$ dependence of $|\psi|^2$.  
The other important observation is that a more poorly screened 
the Coulomb interaction (larger $\kappa_0$), generally tends to 
stabilize the incommensurate phase.  This can be seen from the 
fact that, the final term in Eq. \eqref{FI} is generally negative.  
Since $T_I$ is the first temperature at which the quadratic term 
(in powers of $\psi$) in Eq. \eqref{FI} becomes negative, it is manifest that  
$T_I$ is an increasing function of $\kappa_0$.  However, even for finite $\psi$, 
this term is negative so long as $|\tilde\psi |^2 < 1+2A$, and so 
it generally tends to favor the incommensurate over the commensurate phase, as well. 

This final observation may be significant for understanding the remarkable stability of 
weakly incommensurate CDW states.  When the Coulomb interaction is fully screeened 
($\kappa_0=0$), the commensurate to incommensurate transition occurs when the gain in 
commensurability energy, equals the loss in elastic energy, 
$2V_{c}|\psi|^4=(1/2)K\delta^2|\psi|^2$.  Since both $V_{c}$ and $K$ are typically 
determined by the electronic structure,  unless either $\delta$ is large or $\psi$ 
is small (which typically means that $T$ is close to $T_I$), we expect universally 
to see only commensurate states.  However, where the Coulomb interactions are poorly 
screened ($\kappa_0$ large), there is an additional energetic cost 
$\sim (1/2)\kappa_0(\Delta\bar\rho)^2$ for the commensurate state, 
which could stabilize the incommensurate state to low temperatures, 
even if $\delta \ll 1$. 

In the above we have kept only terms to order $\psi^4$ and lowest order 
in the density fluctuations.  Near a continuous transition between a disordered 
phase and an incommensurate CDW, this is justified, at least at mean-field level, 
on the basis of the small magnitude of the order parameter.  
Since the incommensurate to commensurate transition occurs only when the order 
parameter exceeds a critical magnitude, the above treatment is only valid when 
$\delta$ is small enough.  More generally, at temperatures well below $T_I$, the 
low energy physics, and the commensurate to incommensurate transition in particular, 
should be treated in terms of a phase-only model.  Thus, as long as we are comfortably 
below the mean-field transition temperature, we can integrate out the amplitude modes 
of the CDW order parameters (and the fluctuations of $\rho_N$, as well), and 
concentrate on the low-energy physics of the phase degrees of freedom. Thus, we set 
$\psi_{jn}(\vec r) = |\psi_{j}|\exp[i\theta_{jn}(\vec r)]$.  
To begin with, again consider the stripe case, in which $\psi_{1}=\psi$ and $\psi_{2}=0$;  
then, the effective free energy for the phase degrees of freedom is
\begin{widetext}
%\begin{equation}
\begin{align}
{\mathcal F}_{\rm eff}[\theta] = &
 \sum_{n} \int d^2 x \left\{\frac{\kappa_{\parallel}}{2}
 \left(\partial_x\theta_n - \delta_0\right)^2 +
\frac{\kappa_{\perp}}{2} \left(\partial_y\theta_n\right)^2- U\cos[4\theta_n]\right\} \nonumber\\
&+\frac{g}{2} \sum_{n,n^{\prime}} \int d^2 x \ d^2 x^{\prime} 
\left(\partial_x\theta_n-\delta^{\prime}_0\right) 
\widetilde V(\vec x-\vec x^{\;\prime}, n-n^{\prime}) 
\left(\partial_{x^{\prime}}\theta_n-\delta^{\prime}_0\right) 
%&
\label{Feff-theta}
\end{align}

%\end{equation}
\end{widetext}
where $\widetilde V(\vec x-\vec x^{\;\prime},n-n^\prime)$ is the screened Coulomb 
interaction potential. 
%Here $g$ is an effective coupling constant which scales like 
%$g \propto (e^2/\varepsilon)\;\psi^4$. 
In what follows we will find it convenient to use a form of the screened 
interaction potential which is the solution of 
\begin{eqnarray}
\left\{
-\left(a_z^{-2} \triangle+\nabla^2\right)+\xi_s^{-2}
\right\} &&\!\!\!\!
\widetilde V\left(\mathbf{x}-\mathbf{x}^\prime; n-n^\prime\right)=
\nonumber \\
&&\;\;\;\;\;\;a_z^{-1} \delta_{n,n^\prime} \delta(\mathbf{x}-\mathbf{x}^\prime)
\nonumber \\
&&
\label{V}
\end{eqnarray}
where we have set $\triangle f(n)\equiv f(n+1)+f(n-1)-2f(n)$, and $\xi_s$ is the 
screening length .  

The expression for ${\mathcal F}_{\rm eff}[\theta]$, the effective free energy 
for the phase degrees of freedom, Eq.\eqref{Feff-theta}, can be derived from 
the Landau-Ginzburg theory above, Eq.\eqref{LGW}, which results in expressions 
for the effective stiffness constants  $\kappa_{\alpha}$, the commensurability 
potential, $U$, and the second incommensurability, 
%$\delta_0$ and 
$\delta^{\prime}_0$, in terms of the parameters of the Landau-Ginzburg model.  
The only important aspect of this for our purposes is that $U\propto \psi^4$, 
$\kappa_{\alpha} \propto K_{\alpha}\psi^2$, and $g \propto (e^2/\varepsilon)\psi^4$.  
It is also important to note that in the neighborhood of a surface, 
the parameters in ${\mathcal F}_{\rm eff}[\theta]$ inherit layer index, $n$, dependence.
% both from any explicit $n$ dependence of the parameters in the non-linear sigma model, 
%and from the induced $n$ %dependence of the magnitude of the order parameter, $\psi$.

We conclude with a final observation concerning the checkerboard phase.  
The Landau-Ginzburg free energy in Eq.\eqref{LGW}
 has no direct coupling between the phase of $\psi_1$ and that of $\psi_2$.  
 Thus, to this order, 
 ${\mathcal F}_{\rm eff}[\theta]$ for the checkerboard phase is simply two, 
 totally independent copies of the above effective free energy.  
 In the incommensurate phase, this reflects an exact symmetry 
 - the origin of the two components of the CDW can be shifted relative to each 
 other with no cost in energy.  This has implications for the fluctuation spectrum 
 of an incommensurate checkerboard phase.  (For the commensurate phase, the phase 
 is locked to the underlying crystalline lattice, in any case, so although there 
 are higher order couplings that link the two phases, they are not important. )

\section{The nature of the extraordinary Commensurate-Incommensurate transition}
\label{sec:extra}

There are two cases of interest. In the first case, we can envisage a situation in 
which the coefficient $r$ for the surface layer is different than in the bulk, 
and such that the surface orders 
%in an incommensurate CDW state 
while the bulk remains disordered. This is the direct analog of the 
``ordinary extraordinary'' surface phase transition, which has been well studied 
in magnetic systems \cite{extraordinary}. (In the next section we give a brief 
discussion of the microscopic  physics that can lead to this situation.) 
A mean-field state of this type has the form 
$\psi_n (\vec x) \sim A_n \exp[i(1+\delta_n)\vec Q \cdot \vec x]$ 
%where $\vec Q$ is in the $xy$ plane, and  
where $\lim_{n \to \infty} A_n=0$ exponentially fast 
(on a length scale of the order of the bulk correlation length.) 
The only difference from the ordinary case is that, because of the Coulomb interaction, 
the incommensurability, $\delta_n$, varies from plane to plane, approaching an asymptotic 
value $\lim_{n\to\infty} \delta_n=\delta_0$.  It is straightforward to construct this state 
using the Landau-Ginzburg theory of Eq.\eqref{LGW}.

%We will now show that, due to the competition between Coulomb interactions 
%(even with a finite screening length) and the pinning potential,
% there is an {\em extraordinary} commensurate-incommensurate transition in this system 
%-- that is, for a system with an open surface, the system is in an inhomogeneous state 
%with a charge distribution commensurate at the surface while the bulk remains 
%incommensurate. Since this physics holds at temperatures far below the bulk 
%disordered-incommensurate transition we will work with the phase only model 
%of Eq.\eqref{Feff-theta}. 
 
 The second case of interest, which is the focus of this section,  
 does not have an obvious analog in surface phase transitions 
 in magnetic systems (although it may happen in incommensurate 
 spin-density-wave systems as well.) Here we imagine that the temperature 
 is well below the critical temperature for  bulk incommensurate order so  
 both in the bulk and at the surface, the magnitude of the order parameter is 
 large and essentially fixed. We can now ask if it is possible for the 
 commensurate-incommensurate transition to occur at the surface at a {\em higher} 
 critical temperature than in the bulk. We can discuss the physics of this state 
 in the simpler phase-only model of Eq.\eqref{Feff-theta}.  
 To simplify the analysis, in what follows we will focus on the special case 
 $\delta^{\prime}_0=\delta_0=\delta$ as this does not change the qualitative 
 properties of the solutions, and it greatly simplifies the algebra.
 % (we know from the discussion given above that $\delta$ lies between $\delta_0$ 
 %and $\delta^{\prime}$).
 
We construct this inhomogeneous state as follows: We first note that for the bulk 
homogeneous commensurate state $\theta_n(\vec x)=0$ everywhere, while for the bulk 
homogeneous incommensurate state $\theta_n(\vec x)=\delta \; x$ everywhere 
(assuming stripe order perpendicular to the $x$ axis).  
As in the previous section, we have thus neglected the physics of near-commensurability, 
{\it i.e.} discommensurations, within a given plane - including this physics greatly 
complicates the analysis without changing the qualitative conclusions.  
We will construct an inhomogeneous state which is commensurate at the surface, 
and hence we set $\theta_0(\vec x)=0$, but incommensurate everywhere else, 
$\partial_x\theta_n(\vec x) \neq 0$ for $n=1,2,\ldots$.

 In short, we need to  study the circumstances under which there exists an approximate 
 mean-field state (which minimizes ${\cal F}_{\rm eff}[\theta]$)  in  which (by assumption) 
 the   phase field on each plane has a constant gradient, 
 $\partial_x \theta_n(\vec x)\equiv f_n+\delta$  (discommensuration-free), 
 but which varies from plane to plane.
 % so that asymptotically it obeys the bulk solution. 
This solution must satisfy the boundary conditions $f_0=-\delta$, {\it i.e.\/} 
commensurate at the $n=0$ layer, and  $\lim_{n\to \infty} f_n=0$, {\it i.e.\/} 
the bulk incommensurate state. It should be stressed that while approximate, 
these configurations are upper bounds to the actual non-linear solutions. 
%In particular the state thus constructed is a good approximation provided 
%the correct ground state has few and far separated discommensurations. 
%This is correct if the incommensuration $\delta$ is sufficiently small 
%and if the length scale over which the state is inhomogeneous is long.  
The effective free energy (per unit area) for configurations of this type 
is readily found from Eq.\eqref{Feff-theta} to be
\begin{equation}
{\mathcal F}_{\rm eff}[f]=\frac{\kappa_\parallel}{2} \sum_{n=0}^\infty f_n^2-
\sum_{n=0}^\infty U_n+\frac{g}{2}\sum_{n,m=0}^\infty f_n f_m G_{1D}(n-m)
\label{Feff-1D}
\end{equation}
where $G_{\rm 1D}(n-m)$ is given by
\begin{equation}
G_{\rm 1D}(n-m)=\left(\frac{\mu}{\mu^2-1}\right)\; \mu^{-|n-m|}
\label{G1D}
\end{equation}
with
\begin{equation}
\mu=\frac{1}{2} \left\{ 2+(a/\xi_s)^2+ \left[(2+(a/\xi_s)^2-4\right]^{1/2} \right\}
\label{mu}
\end{equation}
For configurations of this type, the contribution of the pinning potential vanishes for 
all $n \geq 1$ (as they are incommensurate), while on the commensurate surface layer, 
$n=0$, it contributes with the surface value of the pinning potential, $U_s$. 
Hence, $U_n=U_s \delta_{n,0}$.

The unique configuration $f_n$ which minimizes ${\mathcal F}_{\rm eff}[f]$ and 
satisfies the boundary conditions is
%, $f_0=-\delta$ and $\lim _{n \to \infty} f_n=0$, is given by
\begin{equation}
f_n=-\delta \; \left(1-\frac{\gamma}{\mu}\right)\; \gamma^{-n}
\label{solution}
\end{equation}
where $\gamma$ satisfies the identity
\begin{equation}
\left(\gamma+\frac{1}{\gamma}\right)=\left(\mu+\frac{1}{\mu}\right)+
\frac{g}{\kappa_\parallel}
\label{gamma}
\end{equation}
Since $\gamma>\mu>1$,  the solution decays to the bulk value on a scale 
$a/\log \gamma$ shorter than the length scale of the screened interaction. 
The free energy (per unit area) of this solution is 
\begin{equation}
{\mathcal F}= U_{s}^{\rm crit}-U_{s}, \ \ \ \ U_s^{\rm crit}
\equiv \frac{1}{2} \delta^2  g \left(\frac{\gamma}{\mu} \right) 
\left( \frac{g}{\kappa_\parallel}\right).  
\label{energy-surface}
\end{equation}
Thus, if the surface value of the pinning potential $U_{s}$ is greater than 
$U_{s}^{\rm crit}$, the uniform incommensurate CDW state is unstable at the surface.
% and the stable state is this inhomogeneous state, commensurate at the surface but 
%incommensurate in the bulk, and there is indeed an extraordinary commensurate-incommensurate 
%transition. Also 
Moreover, since
%, as seen from Eq.\eqref{energy-surface}, 
$U^{\rm crit} \propto \delta^2$, for the case of a system which is only {\em weakly} 
incommensurate in the bulk, it requires a very small value of the surface commensurability 
coupling to stabilize a commensurate state, there. 

This is not quite the whole story.  The same analysis can be applied to look for a bulk, 
inhomogeneous state, in which a  periodic arrangement of (possibly far separated) planes 
are commensurate, while the intervening planes have incommensurabilities that can be 
obtained in similar fashion by minimizing Eq. \ref{Feff-1D}.  An upper- bound to the 
energy of such an inhomogeneous bulk state is given by using the surface solution we 
have just described, but with an arbitrary layer in the bulk taken to be the layer 
which is commensurate - it is in fact possible to do somewhat better than this.  
Thus, there is a critical value, $U^{\rm crit} < U_s^{\rm crit}$, such that when 
$U > U^{\rm crit}$, there is a bulk instability of the uniform state.  
To find a circumstance in which there is a surface instability, but no bulk instability, 
it is necessary that $U$ is larger in the surface layer than in the bulk, so that
$U_s > U_s^{\rm crit} > U^{\rm crit} > U$.   As we will see in the next section, this is 
possible when a surface phonon results in an enhanced magnitude of the CDW in the surface 
layer, and hence an enhanced tendency toward commensurability.

% Discuss fluctuation effects.  (We still need to think through the issue of fluctuations 
%and the melting of checkerboard states, and/or 
% contact Subir on this, to figure out 
%whether there is anything smart to say about stronger order favoring 
%checkerboards and weaker 
% order favoring stripes, if this is true.)

\section{A possible phonon ``mechanism'' of an extraordinary transition }
\label{sec:phonons}

In this section, we discuss a simple model, motivated by the structure of NaCCOC, 
in which an electron-phonon coupling can lead to an enhancement of the surface tendency to 
CDW order and commensurability.  Of course, there are many possible surface effects, 
so this discussion should be taken as illustrative rather than ``realistic.''

Each Cu site in NaCCOC sits at the center of an octahedron, with the apical (out of plane) 
sites occupied by a Cl, 
instead of the O that appears there in LSCO.   It is known that NaCCOC cleaves such that the 
surface layer is a  
Ca-O layer, so the topmost Cl is exposed at the surface.  It is thus highly plausible 
that the motion of the apical Cl 
is less constrained due to the absence of material above it.  
It is known\cite{apical-lsco}, moreover, that the motion of the apical atom is strongly 
coupled to the charge 
density in the  copper-oxide plane  --
 the apical O moves $ 0.013$ \AA\ closer to the Cu in optimally doped LSCO than in 
 undoped {\LCO} (in YBCO, the apical O displacement is even larger\cite{apical-ybco} 
 $\sim 0.15$ \AA .)

We therefore consider the effect of coupling to an Einstein phonon corresponding to 
the motion of the local charge density in the plane,
\be
H_{\rm el-ph} = \frac{P^2}{2M}+ \frac{1}{2}M \omega_0^2\; X^2  + 
\lambda \; X \; [\rho(\vec x)-\bar\rho].
\ee
Here $P$ and $X$ are the phonon momentum and displacement, $\omega_0$ 
is the phonon frequency, $\lambda$ is the electron-phonon coupling, 
and $\rho(\vec x)$ is the local electron density.  
If we further assume that the phonon is ``fast'' 
($\omega_0$ large compared to the frequency scales of interest), 
we can integrate it out to obtain an effective attraction
\be
H_{\rm eff} = -\frac{\lambda^2}{2M\omega_0^2} \; [\rho(\vec x) - \bar \rho]^2
\ee
when we further substitute the expression in Eq. \eqref{density} for $\rho$ 
in terms of the CDW order parameter, we see that the electron-phonon coupling 
leads to a renormalization of the various parameters that enter the Landau-Ginzburg 
free energy functional, but most importantly, it leads to a negative additive shift of $r$
\be
r \to r - 2\; \frac{\lambda^2}{2M\omega_0^2}
\ee
or equivalently, to an upward renormalization of the mean-field ordering temperature.  
If the surface phonon is softer than in the bulk ({\it i.e.\/} the elastic constant 
$k=M\omega_0^2$ is smaller on the surface), as the above discussion suggests, then 
this renormalization is larger at the surface than in the bulk.  This implies both 
that the ordering temperature at the surface is enhanced, and at a given temperature, 
the magnitude of the order parameter is increased, thereby increasing the chance of a 
commensurate lock-in.  

It is important to note that a large shift in the CDW ordering can occur for rather 
small displacements of the apical Cl positions.  To make a dimensional estimate of 
the expected magnitude of this displacement, we note that the contribution of this 
interaction to the condensation energy is $E_{el-ph} \sim  M \omega_0^2 (\Delta X)^2$;  
this must be less than or equal to the full condensation energy, and hence we can make 
an upper-bound estimate,  $E_{el-ph}\sim \rho(E_F)T_c^2$. If we further crudely estimate 
that $M \omega_0^2\sim E_F/a^2$, we find that
$\Delta X \sim a (T_c/E_F)$, which is generally small.  
%For NaCCOC, this means $\Delta x \sim 10^{-2}\AA\ $.

\section{ Experiments in the cuprates}
\label{sec:NaCCOC}

Dramatic evidence of CDW order in a high temperature superconductor was recently 
obtained from low temperature ($T=4$K) STS experiments of Hanaguri and coworkers
 \cite{hanaguri-davis-oxychloride-04} on NaCCOC
%$Ca_{2-x}Na_xCuO_2Cl_2$ 
with $x=0.08, 0.10$, and $0.12$.  
All of these doping levels are less than the optimal value $x=0.15$, where $T_c$ 
reaches $20 K$;  for $x=0.08$ there is no bulk superconductivity at all. 
For all doping levels, the tunneling conductances exhibit a pseudogap structure for 
energies less than about 
$150 meV$, but no sign of the coherence peaks at lower energies that have been 
identified with the superconducting 
gap in earlier STS studies of BSCCO\cite{howald-prb-03,fisher}. 
Although high-energy topographic maps do not exhibit any periodic modulations 
other than those associated with the underlying crystal structure, 
within the pseudogap, STS reveals large amplitude (order 1) spatial modulations  of 
the local density of 
states (LDOS) with a checkerboard pattern.  
Fourier transforms of
the STS maps reveal peaks corresponding to a commensurate modulation 
 with a $4a_0\times 4a_0$ periodicity and peak widths of order of one tenth of the 
 Brillouin Zone dimensions. 
 
The large amplitude LDOS modulations observed at low energies are reminiscent of those 
seen in conventional CDW systems such as the
dichalcogenides \cite{Coleman}. In those systems, the CDW also shows up in the 
topographic maps. 
Correspondingly, in NaCCOC, one might have expected a signal from the height modulations of 
the 
surface Cl atoms to show up in
the high-energy topographs as a result of CDW-induced atomic displacements. 
However, according to  the estimates in the previous section, for a range of plausible 
$T_c/T_F \sim 10^{-1} - 10^{-2}$, the expected magnitude  of these displacements is in the 
range 
$0.1 - 0.01$ \AA , 
which probably would be undetectable on the grey-scale maps in 
Ref. \onlinecite{hanaguri-davis-oxychloride-04}.   
Another difference with conventional CDW behavior is the doping independence of the 
ordering vector in NaCCOC. 
In conventional CDW systems, the CDW ordering vector changes as the location of the 
Fermi surface changes.  
Finally, whereas the STS modulations seen in the dichalcogenides are highly coherent, 
the correlation length of about $10$
 lattice constants found in NaCCOC is less than 3 periods, making any definitive 
 statements about the character 
 of the order difficult.

To determine whether the observed modulations are indicative of bulk CDW order, 
we need to consider what other signatures of CDW order would be expected.  
There are presently no high resolution  STS studies on NaCCOC at higher temperatures, 
so little is directly known about the thermal evolution of the checkerboard order.  
X-ray or neutron diffraction are the traditional sources of  definitive evidence for  
charge order; 
the in-plane components of
the wavevector are well-defined by the STS experiments, but the body-centered tetragonal 
coordination of the the 
Cu atoms gives
little reason to suspect significant coherence between planes. 
As a result, peaks in the scattering intensity should form rods at wave-vectors
$\mathbf{Q}=(2\pi/a)(1/4,0,\ell)$. However, we are not aware of any reports of any such 
diffraction peaks in NaCCOC. 

%***
A phase transition to a density wave
state is expected to affect the resistivity, in general, by removing some part of the 
Fermi surface (FS) and by modifying scattering
rates on the remaining FS. 
This expectation is generally realized in conventional 
CDW systems.  
The effect on the resistivity is especially strong in cases in which the CDW order 
is sufficiently strong 
that there is a low temperature commensurate lock-in.
%, although whether the larger electromagnetic signature is associated 
%with the initial CDW 
%transition or 
%the commensurate lock-in.
%seems to be species dependent. 
%EF
In contrast,
the temperature dependence of the 
resistivity of NaCCOC\cite{waku-transport-oxychloride} does not exhibit
 any distinct features we can associate
with a phase transition below $300 K$. Moreover, the magnitude of the in-plane 
resistivities are close to what is reported\cite{Ando2004} for 
 LSCO and YBCO, where there is little or no static CDW order.
 
 %SB
 More evidence against the existence of a bulk CDW comes from an examination of the 
 electromagnetic response. 
%Furthermore, 
Optical conductivities of conventional CDW systems show a shift 
of 
%SB
(typically most of the)
oscillator strength to energies above the 
single-particle gap, whereas there is no evidence for such a shift in 
NaCCOC \cite{waku-transport-oxychloride}.   
If there were bulk CDW order in NaCCOC with an ordering temperature below $300 K$, 
surely it would have produced a detectable feature in the electromagnetic response.  
If the ordering temperature were above $300 K$,
surely it would have produced a large quantitative change of the resistivity.  
(Note:  it is usual in
CDW systems that  the ratio of the CDW gap to $T_c$ is large, $2\Delta/kT_c \sim 10$, 
see Ref. \onlinecite{GrunerRMP}.  For a gap size of $\Delta\sim 150\; meV$, 
one might therefore expect a bulk ordering temperature  of around $300 K$.)  

The one caveat on this argument is that, already in earlier studies of the onset of 
stripe order in LNSCO, 
it was observed that while there is a characteristic signature in the temperature 
dependence of the resistivity \cite{ich} 
and a ``localization'' like suppression \cite{Dumm} of the low frequency optical 
conductivity associated 
with the onset of charge order, these features are considerably more muted than in 
conventional CDW systems.  
Presumably, the difference reflects the different origins of the charge order. 
In conventional CDW systems, the ordering is at least loosely associated with 
Fermi surface nesting, and hence the fluctuations above T$_c$ cause singular scattering 
of the 
quasi-particles across these 
nested portions, and a gap is opened on the  Fermi surface in the low temperature phase.  
In contrast, in the cuprates, 
the charge ordering is a strong coupling effect \cite{rmp}, not directly associated with 
any identifiable Fermi surface 
nesting vector, 
and hence the effect of the onset of order on the low energy quasiparticle dynamics is 
much more subtle.  
(The fact that the quasiparticles are always relatively short-lived, in any case, may 
exacerbate this effect.)  
It is thus possible that in NaCCOC, the effect of the CDW ordering on the electrodynamics 
is simply so small as 
to have escaped detection.  However, given that the order is commensurate 
(while that in LNSCO is incommensurate
%EF
and weak
), 
and that the charge order in NaCCOC apparently produces a pseudo-gap in which the density 
of states essentially 
vanishes at zero energy, 
we conclude that it is unlikely that this order, if present in the bulk, would not produce 
a detectable signature 
in the electrodynamics.
%***

We conclude that it is likely that the commensurate CDW order in NaCCOC resides only on 
the surface as an 
extraordinary state of the type described in Section \ref{sec:extra}.  This conclusion is 
suggested by the  absence of any evidence of bulk density wave order,  the absence of the 
expected evolution of the
CDW periodicity with doping level, and the absence of superconducting coherence peaks at 
the surface\cite{asymmetry}. 

Is the checkerboard order, then, simply a surface artifact, from which we learn nothing 
about a bulk tendency toward charge order?  
%This is certainly possible.  However, 
We think a more plausible interpretation is that density wave order, already incipient 
in the bulk, is stabilized at the surface, most probably by a softening of a surface 
phonon.  

In the first place, modulations of the low energy LDOS with similar period, although 
somewhat incommensurate 
and much smaller in magnitude than those seen in NaCCOC, have been documented
\cite{davis-vortex-science-02,howald-prb-03,ali-science-04} on BSCCO surfaces, 
with a long correlation length of the order of
$ 80$ \AA , and interpreted (rightly we believe) as being induced by the disorder pinning of 
some form of 
incipient (fluctuating) CDW order.  In the case of BSCCO, the relevant Cu-O planes are not 
exposed on cleaving 
the crystal, 
but are rather buried under a Bi-O layer.  Thus, there is more reason to hope that the surface 
electronic structure 
is 
similar to that in the bulk.  Moreover, clear signatures of the superconducting gap have been 
reported in STS 
studies 
of BSCCO surfaces \cite{fisher}, again suggesting that the bulk electronic structure is well 
preserved at the surface.  
The fact that the periodicity of the observed modulations is similar (although not equal) to 
those in NaCCOC, 
suggests 
that they are related phenomena, and so supports the  notion that they both reflect 
interesting bulk correlations.

Secondly, neutron scattering studies of LSCO and YBCO reveal ubiquitous evidence of 
fluctuating stripe order
\cite{tranquada-nature-95,vicandme,emery-kivelson-tranquada-pnas-99,lake-nature-02,
rmp,zaanenreview,subireview} 
(that is, a strong enhancement of the dynamical structure factor at low 
$\omega$ and at the characteristic, stripe ordering wave-vector, $\vec q_{\rm stripe}[x]$, 
smoothly dependent 
on the doping fraction $x$) and a weak (possibly extrinsically stabilized) tendency toward 
static spin-stripe order.  
In LSCO, $\vec q_{\rm stripe } [ x ] $ is incommensurate in the relevant range of doping, a 
fact which can be inferred 
from its absolute magnitude, its continuous $x$ dependence, and the small rotation from the 
Cu-O direction 
induced by the weak orthorhombicity of  LSCO \cite{birgeneau}.  Nevertheless, in 
the relevant range of doping, 
the implied periodicity is only slightly greater than $4a_0$.  Similar statements 
apply to the bulk properties of YBCO, 
with the difference that there is still less tendency toward static CDW order.
The fact that the structures seen in diffraction correspond to unidirectional 
density wave order (stripes) while the checkerboards preserve the point-group 
symmetry of the tetragonal lattice, would seem to differentiate these two phenomena.  
However,  it is a rather subtle energy which leads to the selection of stripe 
vs checkerboard order, 
while the basic tendency to charge order and the characteristic ordering 
wave-vector is considerably more robust.  
This is certainly the case, as stressed previously\cite{rmp,vicandme}, 
when the charge order results from a form of Coulomb frustrated phase separation.

Finally, we turn to speculation concerning possible surface phenomena in BSCCO.  
BSCCO is highly micaceous, and 
the top bilayer is concealed below a Bi-O layer.  Thus, there is good reason to 
think that the properties of this top 
bilayer resemble the properties of the bulk.  However, there is one effect, which 
even if weak, may be important since 
it breaks a symmetry of the bulk.  In the bulk, the two layers of the bilayer are 
equivalent, so that in the absence of 
bilayer splitting there should be a single, doubly degenerate band.  
At the surface, this symmetry is broken, in that 
the upper layer is closer to the surface.  Thus means that, even absent 
bilayer splitting, there should be two distinct 
bands.  This effect may need to be considered when discussing evidence of bilayer 
splitting from ARPES studies 
of BSCCO\cite{aharon}.\\

\begin{acknowledgments} 
 
We are grateful to Peter Armitage, J.C. S\'eamus Davis, T. Hanaguri, Aharon Kapitulnik, 
Dung-Hai-Lee, Subir Sachdev, John Tranquada, Ali Yazdani and Shoucheng Zhang for many 
illuminating conversations. We are also grateful to Adrian del Maestro and Subir Sachdev 
for communicating their unpublished results with us. This work was
supported in part by the National Science Foundation through the
grants NSF DMR-04-42537 at the University of Illinois (EF), NSF DMR-04-21960
at UCLA/Stanford (SAK),
NSF DMR-02-03806 at UCLA (SEB), and NSF PHY-99-07949 at the Kavli Institute for 
Theoretical Physics, UCSB, where EF and SAK where participants at the KITP Program on 
{\em Exotic Order and Criticality in Quantum Matter}. EF and SAK thank KITP Director  
David Gross for his kind hospitality.
 
\end{acknowledgments}


\begin{thebibliography}{99}
\bibitem{asymmetry}
%The STS spectra for NaCCOC  
Not only do the STS spectra of NaCCOC  not show sharp coherence peaks even in doping 
regimes in which it is a bulk superconductor\cite{hanaguri-davis-oxychloride-04}, but 
they also show a marked particle-hole asymmetry down to the lowest bias voltages. 
In contrast, the STS spectra in nearly optimally doped BSCCO show a clear particle-hole 
symmetry below the energy scale of the (well defined) coherence peaks. 
% This behavior is characteristic of superconducting systems. 
On the other hand, in very underdoped BSCCO there is little evidence of coherence paks 
and the STS spectra is similarly asymmetric as in NaCCOC\cite{kyle}.
\end{thebibliography}
\end{document}